\documentclass[11pt,a4paper]{article}
\usepackage{amsmath}
\usepackage{amsthm}
\usepackage{amsfonts}
\theoremstyle{plain}
\theoremstyle{definition}

\theoremstyle{remark}
\newtheorem{rem}{Remark}[section]
\newtheorem{exm}{Example}[section]
\def\link#1#2{
\unitlength=0.5cm
\hbox{
\begin{picture}(1,2)(0,0.5)
\put(0,0){\vector(1,2){0.7}}
\put(1,1.7){\makebox(0,0){\hbox{\small $#2$}}}
\put(0.3,-0.4){\makebox(0,0){\hbox{\small $#1$}}}
\end{picture}
}\unitlength=0.01em}

\def\dlink#1#2#3{
\unitlength=0.5cm
\hbox{
\begin{picture}(1,3)(0,1.35)
\put(0,0){\vector(1,2){0.7}}
\put(0.7,1.5){\vector(-1,2){0.7}}
\put(1,1.5){\makebox(0,0){\hbox{\small\it #2}}}
\put(0.3,-0.2){\makebox(0,0){\hbox{\small\it #1}}}
\put(0.3, 3.0){\makebox(0,0){\hbox{\small\it #3}}}
\end{picture}
}\unitlength=0.01em}

\def\CA{{\cal A}}
\def\id{\hbox{id}}

\def\CM{{\cal M}}

\def\CI{{\cal I}}
\def\CJ{{\cal J}}
\def\beq{\begin{equation}}
\def\eeq{\end{equation}}
\newcommand{\bea}{\begin{eqnarray}}
\newcommand{\eea}{\end{eqnarray}}
\def\G{\Gamma}

\def\tsa{\otimes_{{}_{\cal A}}}

\def\Om#1{\Omega^#1(\CA)}
\def\bu{\omega}
\def\bw{\eta}
\def\bv{\xi}

\def\C{\hbox{\bf C}}
\begin{document}
\parindent 0pt
%%%%%%%%%%%%%%%%%%%%%%%%%%%%%%%%%%%%%%%%%%%%%%%%%%%%%%%%%%%%%%%%%%%%%
%%%%%%%%%%%%%%%%%%%%%          Title page
%%%%%%%%%%%%%%%%%%%%%%%%%%%%%%%%%%%%%%%%%%%%%%%%%%%%%%%%%%%%%%%%%%%%%
\title{Noncommutative Differential Geometry and Connections on Simplicial
Manifolds}
\author{Andrzej Sitarz \\
Johannes--Gutenberg Universit\"at, Institut f\"ur Physik,
D-55099 Mainz, Germany \\
{\em e-mail: sitarz@iphthi.physik.uni-mainz.de} }
%%%%%%%%%%%%%%%%%%%%%%%%%%%%%%%%%%%%%%%%%%%%%%%%%%%
\maketitle
\thispagestyle{myheadings}
\def\rightmark{
\begin{tabular}{r}
MZ-TH/96-08 \\
hep-th/9603121
\end{tabular}
}
%%%%%%%%%%%%%%%%%%%%%%%%%%%%%%%%%%%%%%%%%%%%%%%%%%
\begin{abstract}
For a simplicial manifold we construct the differential geometry
structure and use it to investigate linear connections, metric and
gravity. We discuss and compare three main approaches and calculate
the resulting gravity action functionals.
\end{abstract}
%%%%%%%%%%%%%%%%%%%%%%%%%%%%%%%%%%%%%%%%%%%%%%%%%%%
%%%%%%%%%%%%%%%%%%%%%%%%%%%%%%%%%
\section{Introduction}
%%%%%%%%%%%%%%%%%%%%%%%%%%%%%%%%%
One of the greatest open problems in contemporary theoretical
physics is the quantum theory of gravity. Various attempts
to create a consistent theory, which would give the right
classical limit, have been studied in recent years, including
the perturbative approach, supergravity, superstring theories
as well as simplicial gravity. The lattice approach to the
problem of quantization of gravity has been suggested many years
ago by Regge \cite{Regge} and has been studied extensively
(see \cite{DG-REV} for reviews and references) since then.

Since there are no fully successful approaches to the
quantum theory of general relativity on continuous spaces,
which use the path-integral formulation, the approximation
of manifolds by discrete structures is an attractive theory,
from which we can learn something about the properties of gravity.
In this approach the continuous action is seen as a long-distance
effective action at the critical point of the discretized model.

The construction of non-commutative geometry \cite{CONNES} has
opened a new possibility in this area. Generalization of the
tools of differential geometry to the level of algebras has
enabled us to consider discrete structures on the same footing
as continuous manifolds. Therefore, having a discrete lattice
or a simplicial manifold, we may investigate the {\em noncommutative
differential geometry} of such objects. This has already been the
subject of several research papers \cite{SIT,HOI}, which were mostly
devoted to the analysis of structures and gauge theories for such
geometries.

Only recently the topic of non-commutative Riemannian geometry and
linear connections has been investigated for a series of non-standard
geometries \cite{MAD}, including matrix geometries and quantum planes.

In our paper we shall attempt to use these tools to the specific geometry
obtained from a discrete lattice of a simplicial manifold, trying to find
out whether our attempt to derive the structures of Riemannian geometry
and gravity from non-commutative geometry would give a reasonable answer.
Of course, the question, which we would try to answer is whether one may
obtain in this way a prescription for an action, which could correspond to
the classical Einstein-Hilbert action.

Since there is no generally approved definition of the linear connection
for non-commutative geometry we would try to use all possibilities that have
been proposed for various models. Therefore our investigation would
additionally do the job of comparing the proposed definitions, in particular
it would indicate which one can be successfully applied to this geometry.

The paper is organized as follows. First, we give an introduction
of the differential structures on a simplicial manifold, illustrating
it with few simple examples. The detailed construction of the differential
algebra is presented in the appendix.

Following the definition of the metric we present an example of gauge
theory on a simplicial manifold, deriving the Yang-Mills action. Finally
for three definitions of linear connections we present their application
to the discussed geometry.
%%%%%%%%%%%%%%%%%%%%%%%%%%%%%%%%%%%%%%%%%%%%%%%%%%%%%%%%%%%%%%%%%%%%%%%%
\section{Algebras and the First Order Differential Calculus}
Let us assume that we have a {\em simplicial manifold} of dimension
$n$ - i.e., a simplicial complex constructed in such a way that the set of
simplices which contain a vertex (point of a lattice) are homeomorphic to
a $n$-dimensional ball.\footnote{We need this to avoid subtle problems with
considering {\em pseudomanifolds}.} Furthermore, we assume that the simplicial
manifold, to which we should often refer to as {\em triangulation}
is oriented. An orientation of a simplex, shortly speaking,
is an assignment of a sign to an ordered set of its vertices.
The orientation of a $p$-simplex induces, naturally, an orientation of all
its lower-dimensional walls (sub-simplices). Now, we say that a simplicial
manifold is orientable, if one can orient all simplices in such a way, that
if two $p$-simplices share a $p-1$-dimensional wall, then the induced
orientations of this wall are opposite.

In our algebraic approach towards the differential structures
we shall begin with identifying the points of the lattice with
the generators of the commutative algebra of functions. To each point
$p$ of the triangulation we assign a function $p$, which
is $1$ at $p$ and vanishes elsewhere.

Clearly, we have
\begin{equation}
 p s = \delta_{ps} p. \label{a1}
\end{equation}
where $\delta_{ps}$ is the usual Kronecker delta function. For
finite lattices the identity of the algebra could be expressed
as a sum of all the generators
\begin{equation}
1 = \sum_p p.
\end{equation}
%%%%%%%%%%#############################################
%%%%%%%%%%#############################################
The construction of the differential algebra shall be based upon
the geometrical structure of the lattice, and we shall identify links
with one-forms and simplices of the higher dimension with higher order
differential forms. Here, we should only sketch the results, for details
of mathematical construction we refer the reader to the appendix.

Let $\G$ be a free vector space of {\em oriented links} on our
lattice. There exists a natural bimodule structure on $\G$, given
by
\begin{equation}
s \link{p}{q} = \delta_{sp} \link{p}{q}, \;\;\;\;
\link{p}{q} s = \delta_{sq} \link{p}{q}, \label{link1}
\end{equation}
\medskip

The involution operation (complex conjugation) of the algebra $\CA$
extends to $\G$
\begin{equation}
 \left( \link{p}{q} \right)^* = - \link{q}{p}
\label{lstar}
\end{equation}

We identify the above constructed bimodule $\G$ with the one-forms
of the differential algebra over $\CA$. It remains to find an
appropriate linear operator $d: \CA \to \G$ (external derivative).
There exist a unique (up to an automorphism of $\G$) linear map
$d: \CA \to \G$, such that, \vtop{
\begin{itemize}
\item $d$ obeys the Leibniz rule: $d(fg) = (df)g + f(dg)$, $f,g \in \CA$,
    \item ker $d = \mathbb C$
    \item Im $d$ generates $\G$
\end{itemize}}
First, alone from the Leibniz rule we obtain the following restrictions:
\begin{equation}
p dp + dp p = dp, \;\;\;\;\; p ds + dp s = 0, \; p \not= s.
\label{leibniz-1}
\end{equation}
%%%%%%%%%%%%%%%%%%%%
A linear map $\tilde{d}$, which satisfies them must be of the form
%%%%%%%%%%%%%%%%%%%%
\begin{equation}
\tilde{d}(p) = \sum_q C_{pq} \left(
 \link{q}{p} - \link{p}{q} \right),
\end{equation}

where $C_{pq}$ are some arbitrary symmetric constants ($C_{pq}=C_{qp}$).
In particular, if we define a bimodule automorphism $\rho$ of $\G$:

\begin{equation}
\rho\left( \link{p}{q}  \right) \; =  \; C_{pq} \link{p}{q}
\label{exter}
\end{equation}
then $\tilde{d}$ could be written as a composition of $\rho$
with a linear map $d$, $\tilde{d} = \rho \circ d$, where $d$
is as follows:

\begin{equation}
d(p) \; = \; \sum_{q} \left( \link{q}{p} - \link{p}{q} \right)
\label{link2}
\end{equation}
\medskip

Clearly, $d$ is the desired map, which satisfies all our requirements.
Additionally, we find that $d$ is compatible with the previously introduced
$*$-structure on $\Om{1}$ (\ref{lstar}), i.e.:
\begin{equation}
(df)^* = d(f^\star).
\end{equation}

Finally, let us demonstrate that the definition of $d$ is quite natural
and leads, as expected, to the finite difference along oriented links,
being the lattice analogues of partial derivatives. For an arbitrary $\Phi
\in \CA$, $\Phi = \sum_{p} p\, \Phi_p $, we have

\begin{equation}
d \Phi \; = \; \sum_{p,q} \link{p}{q} \; \left( \Phi_q - \Phi_p \right),
\end{equation}
\medskip

\subsection{Higher order differential forms}

The construction of higher order differential forms can be done in many
different ways, which can lead to quite distinct differential algebras.
The point of view we take in our examples is to relate the differential
structure with the geometry of the triangulation. As the details of the
construction are rather formal, we present here only results and the
rules for the multiplication of one-forms and the action of the external
derivative $d$.

First, let us show how to construct higher-order differential forms by
building the product of links. Of course, the multiplication must be done
over $\CA$, therefore the product of two links will vanish unless the
end-point of the first link coincides with the start-point of the second link.
Moreover, the requirement that two-forms correspond to two-dimensional
simplices enforces that the product of the links vanishes unless these
links belong to some $2$-simplex.

For simplicity, we shall denote the product of two links in the following
pictorial way
\begin{equation}
\link{p}{q} \wedge \link{q}{r} \; = \; \dlink{p}{q}{r}
\end{equation}
\medskip

Now, we define the external derivative $d$

\begin{equation}
d\link{p}{q} \; = \; \sum_r \left( \dlink{p}{q}{r} +  \dlink{r}{p}{q}
- \dlink{p}{r}{q} \right). \label{d2}
\end{equation}
\bigskip

Quite easily one may verify that it satisfies the graded Leibniz
rule for the multiplication by functions from the left and right-hand
side

\begin{eqnarray}
d \left( \link{p}{q} \right) = d \left( p \link{p}{q} \right)
= \sum_s \dlink{s}{p}{q} + p\, d  \left(\link{p}{q} \right) \\
d \left( \link{p}{q} \right) = d \left( \link{p}{q} q \right)
= \sum_s \dlink{p}{q}{s} + d  \left(\link{p}{q} \right) q \\
s \not= p: \;\;\; 0 = d \left( s \link{p}{q} \right)
= - \dlink{s}{p}{q} + s\, d  \left(\link{p}{q} \right) \\
s \not= q: \;\;\; = d \left( \link{p}{q} s \right)
= \dlink{p}{q}{s} - d  \left(\link{p}{q} \right) s
\end{eqnarray}
\bigskip

Next, calculate $d^2$,

$$ d^2(p)  \; = \;  \sum_q d \left( \link{q}{p} - \link{p}{q} \right)
\; =  $$
$$ = \sum_{q,r}
\left( \dlink{q}{p}{r} + \dlink{r}{q}{p} - \dlink{q}{r}{p} \right)
- \left( \dlink{p}{q}{r} + \dlink{r}{p}{q} - \dlink{p}{r}{q}
\right), $$

and we can see that the first component in the first bracket cancels with
the second component in the second bracket, and the remaining components
cancel each other within the brackets.

\begin{exm}

To illustrate our construction let us choose a simple piece of
the triangulation of a plane, as drawn on the picture below,

\begin{center}
\unitlength=1cm
\begin{center}
\begin{picture}(5,3)(0,0.5)
\multiput(1,1)(1,0){4}{\circle*{0.1}}
\multiput(1,2)(1,0){4}{\circle*{0.1}}
\multiput(2,0)(1,0){2}{\circle*{0.1}}
\multiput(2,3)(1,0){2}{\circle*{0.1}}
\put(1.2,1.2){\makebox(0.3,0.1){\hbox{\footnotesize (0,1)}}}
\put(1.2,2.2){\makebox(0.3,0.1){\hbox{\footnotesize (0,2)}}}
\put(2.2,0.2){\makebox(0.3,0.1){\hbox{\footnotesize (1,0)}}}
\put(2.2,1.2){\makebox(0.3,0.1){\hbox{\footnotesize (1,1)}}}
\put(2.2,2.2){\makebox(0.3,0.1){\hbox{\footnotesize (1,2)}}}
\put(2.2,3.2){\makebox(0.3,0.1){\hbox{\footnotesize (1,3)}}}
\put(3.2,0.2){\makebox(0.3,0.1){\hbox{\footnotesize (2,0)}}}
\put(3.2,1.2){\makebox(0.3,0.1){\hbox{\footnotesize (2,1)}}}
\put(3.2,2.2){\makebox(0.3,0.1){\hbox{\footnotesize (2,2)}}}
\put(3.2,3.2){\makebox(0.3,0.1){\hbox{\footnotesize (2,3)}}}
\put(4.2,1.2){\makebox(0.3,0.1){\hbox{\footnotesize (3,1)}}}
\put(4.2,2.2){\makebox(0.3,0.1){\hbox{\footnotesize (3,2)}}}

\multiput(1,1)(1,0){3}{\line(1,0){1.0}}
\multiput(1,2)(1,0){3}{\line(1,0){1.0}}
%%\multiput(1,0)(0,1){3}{\vector(0,1){1.0}}
\multiput(2,0)(0,1){3}{\line(0,1){1.0}}
\multiput(3,0)(0,1){3}{\line(0,1){1.0}}
%%\multiput(4,0)(0,1){3}{\vector(0,1){1.0}}
%%\put(1,1.1){\makebox(0.3,0.1){\hbox{\small $\tsa$}}}
%%\multiput(0,0.2)(0,0.3){9}{\circle*{0.1}}
%%\put(1,1.5){\makebox(0.3,0.1){\hbox{\small\it #2}}}
\end{picture} \\

\ \\
Fig.1
\end{center}
\unitlength=0.01em

\end{center}

As it is a two-dimensional triangulation of the plane (so that
all points $p,q,r,s$ lie in the same plane) we see that the
following wedge products of links vanish (as well as their
conjugates):

\vspace{0.5cm}
$$
0 = \dlink{p}{q}{r} \;\; = \dlink{q}{r}{p} \;\; = \dlink{r}{p}{q}
$$
\vspace{0.5cm}

The rest, i.e., products of links that are (different) sides of the
triangle of the lattice are linearly independent. Had the figure
shown the surface of a tetrahedron then $p,q,r$ would be in one
triangle and the above products of links would not vanish.
\end{exm}

Finally, let us follow another illustrating calculation, for an
arbitrary one form,
$$ \omega = \sum_{p,q} \link{p}{q} \omega_{pq} $$
\medskip
calculate $d\omega$,
\begin{equation}
d\omega = \sum_{p,q,r} \omega_{pq} \left( \dlink{r}{p}{q}
+ \dlink{p}{q}{r} - \dlink{p}{r}{q} \right).
\end{equation}
\bigskip

After rearranging the order of components of this sum we obtain
the following convenient expression

\vskip -1cm
\begin{equation}
d\omega = \sum_{p,q,r} \dlink{p}{q}{r} \left(
\omega_{pq} + \omega_{qr} - \omega_{pr} \right).
\end{equation}
\medskip

\begin{rem}
Of course, the cohomology of the complex defined by the sequence:
$$ \CA \xrightarrow{d} \Om{1} \xrightarrow{d} \Om{2}$$
depends on the topology of the triangulated manifold.
\end{rem}

For the products of more than two links we have an additional feature,
which we must take into account. There might exist different (a priori)
products with the same start- and end-points. As we wish that this situation
does not happen (see appendix for motivation and formal construction) we
should identify the corresponding products up to a sign, which originates
from the orientation of the simplex. Then every non-vanishing product
of $n$ links could be written (in a symbolic way) as a $n$-dimensional
simplex multiplied by the start-point (from the left) and the end-point
(from the right). Below we present the example of a
three-dimensional triangulation.

\begin{exm}
We have two possibilities for the products of three links in
a simplex, which have the same start- and end-points. The
relation between them follows from the above description:

\unitlength=1cm
\begin{center}
\
\hbox{
$-$
\begin{picture}(2,2)(0,0.5)
\put(0,0){\vector(1,0){1.5}} %%% (0.1,0){15}{\circle*{0.03}}
\multiput(0,0)(0.1,0.1){12}{\circle*{0.03}}
\multiput(0,0)(0.1,0.06){25}{\circle*{0.03}}
\put(1.5,0){\vector(-1,4){0.3}} %%%% (-0.025,0.1){13}{\circle*{0.03}}
\multiput(1.5,0)(0.1,0.15){10}{\circle*{0.03}}
\put(1.2,1.2){\vector(4,1){1.2}} %%%%(0.1,0.02){12}{\circle*{0.03}}

\end{picture}

\quad = \quad

$\displaystyle  p \left(
\hbox{
\begin{picture}(2.5,1.4)(0,0.5)
\put(-0.2,-0.2){\makebox{$p$}}
\put(2.4,1.5){\makebox{$q$}}
\multiput(0,0)(0.1,0){15}{\circle*{0.03}}
\multiput(0,0)(0.1,0.1){12}{\circle*{0.03}}
\multiput(0,0)(0.1,0.06){25}{\circle*{0.03}}
\multiput(1.5,0)(-0.025,0.1){13}{\circle*{0.03}}
\multiput(1.5,0)(0.1,0.15){10}{\circle*{0.03}}
\multiput(1.2,1.2)(0.1,0.02){12}{\circle*{0.03}}
\end{picture}
}
\right) q $

\quad = \quad

$+$
\begin{picture}(2,2)(0,0.5)
\multiput(0,0)(0.1,0){15}{\circle*{0.03}}
\put(0,0){\vector(1,1){1.2}} %%%%%{\circle*{0.03}}
\multiput(0,0)(0.1,0.06){25}{\circle*{0.03}}
\put(1.2,1.2){\vector(1,-4){0.3}} %%%%% {\circle*{0.03}}
\put(1.5,0){\vector(2,3){1}} %%%%%%%(0.1,0.15){10}{\circle*{0.03}}
\multiput(1.2,1.2)(0.1,0.02){12}{\circle*{0.03}}

\end{picture}
}
\
\end{center}
\unitlength=0.01em
\end{exm}
%%%%%%%%%%%%%%%%%%%%%%%%%%%%%%%%%%%%%%%%%%%%%%%%%%%%%%%%%%%%%%
\section{Metric and distances}

In this section we shall discuss the concept of {\em metric}
and distances on our lattice. The definition we use,
has been introduced and discussed throughout several papers
\cite{SIT}-\cite{MAD}.

Define metric $g$ as a bimodule morphism:
\begin{equation}
g: \Om{1} \tsa \Om{1} \to \CA
\label{defme}
\end{equation}
We shall write $g(\bu,\bv)$ instead of  $g(\bu \tsa \bv)$ to shorten our
notation and make it resembling the classical differential geometry.
We say that the metric is {\em hermitian} if for all
$\bu,\bv \in \Om{1}$:
\begin{equation}
g(\bu,\bv)^* \; = \; g(\bv^*,\bu^*). \label{herm1}
\end{equation}
This guarantees that $g(\bu,\bu^*)$ and $g(\bu^*,\bu)$ are self-adjoint
elements \footnote{Note that the might be, however, different!} of $\CA$.
For the discrete geometry of a triangulation the metric has quite
a simple form and it associates a real number $g_{pq}$ to
every oriented link from $p$ to $q$. This follows easily from
the definition (\ref{defme}) ($g$ being a bimodule morphism):

\begin{equation}
g \left( \link{p}{q} \tsa \link{q}{s} \right)
\; = \; p g_{pq} \delta_{ps},
\label{dismetric}
\end{equation}
where $g_{pq}$ are complex coefficients. However, from the hermicity
(\ref{herm1}) we immediately get that $g_{pq} = \overline{g_{pq}}$.

Observe that the general definition does not require the
symmetry of the metric (whatever it may mean). Moreover, in general we
might have $g_{pq} \not= g_{qp}$, which would tell us that the metric
structure is associated with oriented links.\footnote{An intuitive
picture of that could be that from the point of view of the metric
going from point $p$ to $q$ is not the same as going from $q$ to $p$!}

\begin{exm}

Let $\Psi = \sum_{pq} \link{p}{q} \psi_{pq}$ be an arbitrary one-form, then

\begin{eqnarray}
g(\Psi, \Psi^*) & = & \sum_{p,q} p g_{pq} |\psi_{pq}|^2, \\
g(\Psi^*,\Psi) & = & \sum_{p,q} p g_{pq} |\psi_{qp}|^2,
\end{eqnarray}
and we see that $g(\Psi,\Psi^*)$ cannot be equal to $g(\Psi^*,\Psi)$
for every $\Psi$, no matter what relation between the coefficients of
the metric we choose (apart from  $g \equiv 0$, of course).
\end{exm}

We shall say that $g$ is {\em positive} if $g(\bu,\bu^*)$ and
$g(\bu^*,\bu)$ are positive elements of ${\CA}$ for every $\bu \in \Om{1}$.
The metric $g$ on discrete geometries (as in (\ref{dismetric}))
is positive if and only if every $g_{pq} \geq 0$.

We shall say that the metric is non-degenerate if $g(\bu,\bu^*)=0$
implies $\bu=0$.  Of course, the metric on discrete geometries is
non-degenerate only if $g_{pq} > 0$ for all links $p \to q$.

To solve the apparent problem in the choice between $g(\bu,\bu^*)$ and
$g(\bu^*,\bu)$ let us observe that it is not the algebra-valued expression,
like the metric, but rather a number-valued expression (i.e. the integrated
metric) that is important for physics. Let $\int$ be a trace on the algebra
\CA, such that
$\int 1 = 1$, $\int (a^*) = \left( \int a \right)^*$
and $\int aa^* = 0$ only for $a=0$.

\begin{exm}
For the algebra of functions on the discrete space, such a trace is of the form
\begin{equation}
\int a = \sum_p \mu_p a_p,
\end{equation}
where $a_p = p a$ and $\mu_p >0$ (the latter are arbitrary positive
numbers, such that $\sum_p \mu_p = 1$).
\end{exm}

Suppose now that we demand that the integrated metric is independent of the
order of the arguments, i.e.:
$$ \int g(\psi, \psi^*) = \int g(\psi^*,\psi). $$
Then it follows directly
\begin{equation}
\frac{\mu_p}{\mu_q} = \frac{g_{qp}}{g_{pq}}.
\label{met-mes}
\end{equation}
The above condition could be seen as a compatibility condition between
the measure on the manifold and the metric. Clearly, not every metric
is admissible. Indeed, for any closed loop along the links of the
simplicial manifolds the product of metric coefficients along one
orientation of the loop must be equal to the product along the opposite
orientation. Suppose we have a metric satisfying this condition. Then
there exists a unique measure determined by this metric. The construction
follows directly from the relation (\ref{met-mes}), for some point $x$ let
us fix $\mu_x=1$. Then, for every point $y$ connected with $x$ with a link we
determine that $\mu_y = \frac{g_{xy}}{g_{yx}}$. Then we repeat the procedure
until we reach all the points of the lattice. Of course, since there are
many possible ways to reach a point at some distance from $x$ we must ensure
that this procedure is well-defined. This is the case due to the property of
the metric: for two different ways we see that the end results coincide
because for the loop made out of them, the product of coefficients of
$g$ is the same along either of its orientations. Finally we must rescale
the measure to have $\int 1$ = 1.

We shall see now that these procedure might be used the other way round.
Suppose we have a measure. Then the relation (\ref{met-mes}) fixes the ratio
between $g_{pq}$ and $g_{qp}$ for any two points connected with a link. What
remains a free parameter is a real number, which fixes their values. Now, we
have a clear understanding of an admissible metric for a discrete geometry.
It contains two pieces of information about the manifold: the measure
(which can be interpreted as a positive function on vertices) and the
association of a positive real number to every unoriented) link, we shall
later denote this number $r_{pq}=r_{qp}$, then $g_{pq} = \mu_q r_{pq}$.

%%%%%%%%%%%%%%%%%%%%%%%%%%%%%%%%%%%%%%%%%%%%%%%%%%%%%%%%%%%%%%%%%%%%%%%%
%%%%%%%%%%%%%%%%%%%%%%%%%%%%%%%%%%%%%%%%%%%%%%%%%%%%%%%%%%%%%%%%%%%%%%%%
\medskip

The above definition of the metric leads, however, to some difficulties
if we attempt to extend it to higher-order forms. Of course, it is quite
natural to propose the following extension of $g$ to $\Om{1} \tsa \Om{1}
\tsa \Om{1} \tsa \Om{1}$:
\begin{equation}
g_2(\bu_1 \tsa \bu_2 \tsa \bu_3 \tsa \bu_4) \; = \; g\left( \bu_1
g( \bu_2, \bu_3) \bu_4 \right),
\end{equation}
%%%%%%%%%%%%%%%%%
though, of course, we may use another definition:
%%%%%%%%%%%%%%%%%
\begin{equation}
g_2'(\bu_1 \tsa \bu_2 \tsa \bu_3 \tsa \bu_4) \; = \; g( \bu_1,
\bu_2) g( \bu_3, \bu_4 ).
\end{equation}
%%%%%%%%%%%%%%%%%
Of course, if $g$ is hermitian (\ref{herm1}) so are $g_2$ and $g_2'$,
moreover, for positive $g$ we can easily see that $g_2$ and $g_2'$
become positive.

To use the above definition for the two-forms rather than $\Om{1} \tsa
\Om{1}$ we need to have a bimodule embedding $\rho: \Om{2} \hookrightarrow
\Om{1} \tsa \Om{1}$, such that $\pi \circ \rho = \hbox{id}$, where $\pi$ is
the projection $\Om{1} \tsa \Om{1} \to \Om{2}$. Generally,
there is an ambiguity in defining such map, however, in the
case of the triangulation  we might use the natural embedding $\rho$:
\vskip-0.5cm
\begin{equation}
\rho \left( \dlink{p}{q}{r} \right) = \link{p}{q} \tsa \link{q}{r}.
\end{equation}

Then the definition for the metric on the space of two-forms would
be the following:
\begin{equation}
g(u,v) := g_2( \rho(u), \rho(v) ),
\end{equation}
for every two-forms $u,v$. Let us observe that we are left only with
the possibility of $g_2$ in our definition, as $g \circ \rho \equiv 0$
and hence $g_2'$  vanishes.

As an example of a physical theory, in which we have to use differential
structures and the metric, we shall consider a simple nonabelian
gauge theory on the triangulation.

\begin{exm}
Let us consider an algebra of matrix-valued functions on our lattice with
the gauge group being the group of unitary elements of this algebra. The
gauge potential is a matrix-valued, anti-self-adjoint one-form $A=-A^*$:
\begin{equation}
A \; = \; \sum_{p.q} \link{p}{q} A_{pq}, \;\;\;\; A_{pq} = A_{qp}^\dagger,
\end{equation}

The gauge transformation acts on $A$ in the usual way, the transformation
rule for its coefficients being:
\begin{equation}
A_{pq}' \; = \; U_p^\dagger A_{pq} U_q + U^\dagger_p ( U_q - U_p),
\label{gauge}
\end{equation}
where $U_p$ denotes the value of gauge transformation at point $p$.
The transformation (\ref{gauge}) becomes more transparent if we
introduce a field $\Phi_{pq} = 1 + A_{pq}$:
\begin{equation}
\Phi_{pq}' \; = \; U^\dagger_p \Phi_{pq} U_q.
\end{equation}

The curvature $F = dA + A \wedge A$ is a matrix-valued hermitian two-form,
with coefficients:
\begin{equation}
F = \sum_{p,q} F_{pqr} \dlink{p}{q}{r},
\end{equation}
where
\begin{equation}
F_{pqr} \; = \;  \left( \Phi_{pq} \Phi_{qr} - \Phi_{pr} \right)
\; = \; F_{rqp}^\dagger,
\end{equation}
and its gauge transformation is $F' = U^\dagger F U$:
\begin{equation}
F_{prq}' \; = \; U_p^\dagger F_{prq} U_q.
\end{equation}

The Yang-Mills Lagrangian for the theory would be:
\begin{eqnarray}
{\cal L}_{YM} & = & \hbox{Tr\ } g\left( \rho(F), \rho(F^*) \right)
\; = \nonumber \\
& = &   \hbox{Tr\ } \sum_{p,q,r} p g_{pq} g_{qr} F_{pqr} F_{pqr}^\dagger
\label{YM-reg}
\end{eqnarray}

This Lagrange function is real-valued and gauge invariant. Observe
that if we express it in terms of the field $\Phi$ we get three
components, first one, quartic in $\Phi$:
$$ \sum_{\triangle(p,q,r)}  g_{pq} g_{qr}
 \hbox{Tr\ } \Phi_{pq}^\dagger \Phi_{pq} \Phi_{qr} \Phi_{qr}^\dagger, $$
quadratic in $\Phi$:
$$ \sum_{\triangle(p,q,r)}  g_{pq} g_{qr}
\hbox{Tr\ } \Phi_{pr} \Phi_{pr}^\dagger, $$
and cubic in $\Phi$:
$$ \sum_{\triangle(p,q,r)}  g_{pq} g_{qr}
\hbox{Tr\ } \Phi_{pq} \Phi_{qr} \Phi_{rp} +
\hbox{h.c.} $$

If one restricts the space of all possible connections to {\em unitary}
connections satisfying $\Phi_{pq}^\dagger \Phi_{pq} =1$ (this condition
is gauge invariant and therefore acceptable, also all pure gauge connections
are of this type) the first two components contribute only to constant
terms (volume of space), and only the third one gives us the lattice action
for a restricted pure gauge theory on triangulations.
\end{exm}

%%%%%%%%%%%%%%%%%%%%%%%%%%%%%%%%%%%%%%%%%%%%%%%%%%%%%%%%%%%%%%%%%%%%%%%%%%%%%%%
\section{Linear Connections}

The general concept of Riemannian geometry and linear connections is
still not well understood in the framework of noncommutative geometry.
Several propositions have been used, starting from the idea of
left-linear connections \cite{SIT}, bimodule connections \cite{MAD}
or projective-bimodule connections \cite{CQ}. The problem is related
with the bimodule structure of the space of one-forms, which is the
crucial obstacle for extending the gauge theory formalism of
connections to the Riemannian geometry case.

In this section we shall attempt to give a thorough discussion of
these three main approaches, using the corresponding propositions for
definitions of linear connection, torsion and curvature, for the
example of the simplicial geometry.

Our attempt is to derive the Riemannian geometry of triangulated
manifold, treated as a base space of theory, not as an approximation
of a continuous object.

\subsection{Left-linear connection}

A left-linear connection $\nabla_L$ is a map: $\Om{1} \to \Om{1} \tsa \Om{1}$,
which satisfies:
\begin{equation}
\nabla_L ( a \omega)  = a (\nabla_L \omega) + da \tsa \omega
\end{equation}

This map easily extends to a degree 1 map: $\Om{*} \tsa \Om{1}
\to \Om{*} \tsa \Om{1}$, with the following property:
\begin{equation}
\nabla_L \left( \bu \bw \right) \; = \;
( d\bu)  \bw  + (-1)^{|\bu|} \bu  ( \nabla_L \bw ),
\label{codev-0}
\end{equation}

In particular, we have that
$\nabla_L^2$ is a left-module endomorphism of degree 2 of
\mbox{$\Om{*} \tsa \Om{1}$}.

Let $\pi$ be the projection $\Om{*} \tsa \Om{1} \to \Om{*}$. Then the torsion,
$T = \pi \circ \nabla_L - d \circ \pi$ is a left module homomorphism $\Om{*}
\tsa \Om{1} \to \Om{*}$.

Similarly, as we have done for the left-connection, we may introduce
right-linear connections and the corresponding torsion and curvature.

Let us assume that the differential algebra is a star-algebra (i.e. there
exists a conjugation, which (graded) commutes with $d$).
If $\nabla_L$ is a left-connection then ${\nabla_R} = \pm * \nabla_L *$ is a
right connection, the sign chosen accordingly for the degree of the space of
forms it acts on.

To verify it, let us take $\omega \in \Om{1} \tsa \Om{n}, \rho \in \Om{m}$,
then:
\begin{eqnarray}
\nabla_R (\omega\rho) = (-1)^{m+n+1} (*  \nabla_L *) (\omega \rho) =
(-1)^{m+n+1} * \nabla_L ( \rho^* \omega^*)  \nonumber \\
= (-1)^{m+n+1} \left( (-1)^{m} d \rho^* \omega^*
+ (-1)^{m} \rho^* \nabla_L (\omega^*) \right)^* \nonumber \\
=   (-1)^{n+1} \omega d \rho + (-1)^{n+1} (* \nabla_L  *)(\omega) \rho
\nonumber \\
=  (-1)^{n+1} \omega d \rho + (\nabla_R \omega) \rho \nonumber
\end{eqnarray}
which ends the proof.

Of course, since $\pi$ is a star-homomorphism, we immediately see that
for a torsion-free left connection $\nabla_L$, the conjugated right-linear
connection $\nabla_R$ is a also torsion free.

For the considered examples of differential calculus on triangulations
we shall have:
\begin{equation}
\nabla_L \left( \link{p}{q} \right)
= \sum_s \link{s}{p} \tsa \link{p}{q}
+ \sum_{r,s}  \Gamma^{pq}_{rs} \link{p}{r} \tsa \link{r}{s}
\end{equation}
\medskip
where $\Gamma^{pq}_{rs}$ are arbitrary complex numbers.

The torsion constraint $T=0$ reads:
$$\Gamma^{pq}_{qr} = 1, \;\;\;\;\; \Gamma^{pq}_{rq} = -1,$$
for $p,q,r$ forming a  triangle,
$$\Gamma^{pq}_{rs} = 0$$
for $p,r,s$ forming a triangle, $s \not= q$.

\subsection{Metric compatibility condition}

Having linear connections and the metric we may pose the question about
the notion of compatibility between these two structures.
There is no default answer by now, therefore we should use one of
natural possible suggestions. We say that a left linear connection $\nabla_L$
(and the corresponding right-linear connection $\nabla_R$) are
compatible with the metric $g$ if the following holds for all one-forms
$\omega, \eta$:

\begin{equation}
d g(\omega, \eta) = \tilde{g}_L( \nabla_L\omega, \eta) - \tilde{g}_R(\omega,
\nabla_R\eta), \label{comp}
\end{equation}

where $\tilde{g}$ are the extensions of $g$, for instance:
$$\tilde{g}_L(\omega_1 \tsa \omega_2, \eta)  = \omega_1 g(\omega_2, \eta).$$

Notice that the above constraint includes a lot more that in the
classical case. This is because the left-hand side is by definition
middle-linear (i.e. depends only on $\omega \tsa \eta$, whereas for
the right-hand side this may not hold for general left-linear connections).

Let us verify what is the  outcome of the above introduced
metric-compatibility condition for the simplicial geometry. After
simple calculations we obtain

\begin{equation}
\Gamma^{pq}_{sr} g_{sr} = \overline{\Gamma}^{sr}_{pq} g_{pq},
\label{compat}
\end{equation}
(no summation over repeated indices).

One may verify that this is also a sufficient condition for the
right-hand side of (\ref{comp}) to have the same middle-linearity
property as the left-hand side.

This condition has a very surprising aftermath.
Consider a triangle $p,q,r$, and use the torsion constraint
as well as the metric compatibility condition. Then one gets
\begin{eqnarray}
 g_{rq} & = & g_{pq}, \\
 g_{qr} & = & \overline{\Gamma}^{qr}_{pq} g_{pq},
\end{eqnarray}

and from the first relation (and $g_{pq}=r_{pq} \mu_q$) we immediately
see that all $r_{pq}$ must be equal to each other. Hence the metric is
now restricted only to an overall constant $r$ and the measure $\mu$!.
Furthermore, we obtain an expression for some of the nontrivial symbols
$\Gamma$ (notice that it follows immediately that $\Gamma^{qr}_{pq}$ must
be real).

To summarize, for $p,q,r$ being the vertices of a triangle the torsion
and metric compatibility constraint give us together the following
relations:

\begin{eqnarray*}
\Gamma^{pq}_{qr} = 1 \\
\Gamma^{ps}_{qr} = 0, s \not=q \\
\Gamma^{pq}_{rq} = -1 \\
\Gamma^{qr}_{pq} = \frac{\mu_r}{\mu_q}
\end{eqnarray*}

Having calculated the torsion and metric compatibility conditions we shall
finally attempt to derive the curvature. From the general formula, the
curvature for any left-linear connection on our geometry is:

\begin{eqnarray}
 R(\link{p}{q}) =  \sum_{s,w}
\left( \dlink{w}{s}{p} - \dlink{s}{w}{p} \right) \tsa \link{p}{q} -
\sum_{s,w,t} \Gamma^{pq}_{st} \dlink{w}{p}{s} \tsa \link{s}{t} & & \nonumber \\
& & \nonumber \\
 +  \sum_{s,w,t} \Gamma^{pq}_{st} \left(
\dlink{w}{p}{s} - \dlink{p}{w}{s} \right) \tsa \link{s}{t}
- \sum_{{s,t,w,z}} \Gamma^{pq}_{st} \Gamma^{st}_{wz}
\dlink{p}{s}{w} \tsa \link{w}{z} & &
\nonumber
\end{eqnarray}
\medskip

which reduces to the following compact expression:
\begin{equation}
R( \link{p}{q})  \; = \; - \sum_{s,w,t} \Gamma^{pq}_{st} \dlink{p}{w}{s}
\tsa \link{s}{t} - \sum_{s,t,w,z} \Gamma^{pq}_{st} \Gamma^{st}_{wz}
\dlink{p}{s}{w} \tsa \link{w}{z}
\end{equation}
\medskip

Next, we can introduce a Ricci tensor as a trace of the map $R$:
$$ R ( \link{p}{q} ) \; = \; \sum_{r,s,t} R^{pq}_{prst} \link{p}{r} \wedge
\link{r}{s} \tsa \link{s}{t}, $$
then:
$$Ric = \sum_{p,q,s,t} R^{pq}_{pqst} \link{q}{s} \tsa \link{s}{t}.$$
In our case we get:
\begin{equation}
Ric \; = \; - \sum_{p,q,s,t} \Gamma^{pq}_{st} \link{q}{s} \tsa \link{s}{t}
- \sum_{p,q,t,w,z} \Gamma^{pq}_{qt} \Gamma^{qt}_{wz} \link{q}{w} \tsa
\link{w}{z}.
\end{equation}

The scalar of curvature could now be introduced as a function obtained
by composing $g \circ Ric$, we have $R = \sum_q q R_q$:
\begin{equation}
R_q \; = \; - \sum_{p,w} g_{qw} \Gamma^{pq}_{sq}
- \sum_{p,w,t} g_{qw} \Gamma^{pq}_{qt} \Gamma^{qt}_{wq}.
\end{equation}

Remember that we have fixed only some of the Christoffel symbols and
the rest of them are still arbitrary. Therefore we might divide $R_q$
into two parts, first, which we may already calculate:
\begin{equation}
R_q^0 =  \sum_{(p,w;q)} \mu_w - \sum_{\frac{(p,t;q)}{(w,t;q)}}
\mu_w \frac{\mu_t}{\mu_q} - \sum_{(p,t;q)} \mu_w \frac{\mu_p}{\mu_q}.
\end{equation}

The part, which contains auxiliary fields is as follows:

\begin{equation}
R_q^a = - \sum_{\frac{[p,t;q]}{[w,t;q]}} \mu_w \Gamma^{pq}_{qt}
\Gamma^{qt}_{wq} - \sum_{\frac{(p,t;q)}{[w,t;q]}} \mu_w \Gamma^{qt}_{wq}
- \sum_{\frac{[p,t;q]}{(w,t;q)}} \mu_w \frac{\mu_q}{\mu_t}
\Gamma^{pq}_{qt} - \sum_{(p,q;w)} \mu_w \Gamma^{qw}_{wq}.
\end{equation}

where the brackets indicate that the sum is over point forming certain
triangle and square brackets denote that the sum is over points do not
forming a triangle. Of course, point $q$ is fixed and $p,q,w$ always make
a triangle.

This result is not very promising. Though we have been able to define
the connection, the metric and the relations between them, the connection
still contains some auxiliary fields, which appear also in the resulting
action functional.

\subsection{Cuntz and Quillen linear connections}

Recently Cuntz and Quillen have proposed a definition for linear
connections on projective bimodules. Such connection is a pair of
left and right linear connections (as defined in the previous section),
with the restriction that they are simultaneously a right
(and respectively left) module homomorphism:
\begin{equation}
\nabla^{CQ}_L ( a m b) = a (\nabla_L^{CQ} m) b + da \tsa m b.
\end{equation}

Of course, our considerations that used the conjugation structure
on the bimodule are still valid, therefore we shall still consider
a pair of left and right connections that are related by
$\nabla_R = \pm * \nabla_L *$.

For the case of triangulation geometry it means the following:
\begin{equation}
\nabla_L^{CQ} \link{p}{q} = \sum_s \link{s}{p} \tsa \link{p}{q}
+ \sum_s \Gamma^{pq}_{sq} \link{p}{s} \tsa \link{s}{q}.
\end{equation}

However, the torsion constraint cannot be satisfied in its form
$T = \pi \circ \nabla - d \circ \pi$ and we have to modify it to
satisfy the general properties arising from the Leibniz rule:
\begin{equation}
T = \pi \circ( \nabla_L^{CQ} + \nabla_R^{CQ}) - d \circ \pi.
\end{equation}

We need to modify it and take the connection part of torsion as
$\pi \circ (\nabla_L^{CQ} + \nabla_R^{CQ} )$, which is the
only operator that shares the property of Leibniz rule
for left and right multiplication with $d$.

The torsion $T=0$ constraint then gives us the following
restriction on coefficients $\Gamma$
\begin{equation}
\Gamma^{pq}_{rq} + \overline{\Gamma}^{qp}_{rp} = -1,
\label{CQ-1}
\end{equation}
for all $p,q,r$ forming a triangle.

The metric compatibility condition (\ref{comp}) shall remain the
same
\begin{equation}
\Gamma^{pq}_{sq} g_{sq} = \overline{\Gamma}^{sq}_{pq} g_{pq},
\label{CQ-2}
\end{equation}
however, since we have no further relations there would be no
restriction on the metric itself (in particular $r_{pq}$ are
not fixed in this case). The relations between $\Gamma$ and
the metric have more than one solution. One can easily find that
if $\Gamma$ and $\Gamma'$ satisfy (\ref{CQ-2}) and (\ref{CQ-1})
for a fixed metric, then $\alpha \Gamma + (1-\alpha) \Gamma'$ does
as well, $0 \leq \alpha \leq 1$.

Note that the solutions might not be even real, however. should
at least one $\Gamma^{pq}_{rq}$ be real this would enforce all
$\Gamma$ to be real. Moreover, using simple arguments one can show
that the space of all possible connections compatible with a given
metric is one-dimensional, in particular if one fixes its one elements
$\Gamma^{pq}_{rs}$ then all others could be calculated.

An example of a real connection can be found quite easily,
one verifies that
\begin{equation}
\Gamma^{pq}_{sq} = - \frac{r_{pq}}{r_{pq} + r_{sq}}.
\end{equation}
satisfies (\ref{CQ-1}) and (\ref{CQ-2}).

For the Cuntz-Quillen the curvature $(\nabla^{CQ}_L)^2$
(or equivalently $(\nabla^{CQ}_R)^2$) is a bimodule
morphism $\Omega^1 \to \Omega^2 \tsa \Omega^1$:

$$ R \left( \link{p}{q} \right) = - \sum_{s,w}
\left( \Gamma^{pq}_{sq} +
\Gamma^{pq}_{wq} \Gamma^{wq}_{sq} \right)
\dlink{p}{w}{s} \tsa \link{s}{q} $$

The Einstein-Hilbert action becomes
$${\cal S} = - \sum_{p,q,s} \mu_q g_{qs} \Gamma^{pq}_{sq}, $$
and it appears that, alone, it might not even be real-valued.

What seems to be a bigger problem, is the auxiliary parameter
in the connection. Since the metric determines it only up to
a complex parameter (note that the dependence is not linear),
we would like to have at least the action independent of such
parameter. However the above action does not satisfy our
demand.

In the approach to gravity on the simplicial manifold based on
Cuntz-Quillen connections we have encountered a significant
problem of an auxiliary parameter. Of course, this result is
much better than in the case of left-linear connections, however
unless solved in a satisfactory way, we cannot apply it to
investigate gravity.

\subsection{Bimodule connections}

Finally we shall discuss another proposition for the construction of
linear connections in the framework of noncommutative geometry,
which uses concepts of the generalized symmetry operator $\sigma$
\cite{MAD}. Let us remind the basic assumptions. We assume that
there exists a bimodule automorphism of $\Om{1} \tsa \Om{1}$ such
that $\pi \circ ( 1 + \sigma) \equiv 0$ and we define $\nabla$ to
be a {\em bimodule connection} if $\nabla$ is a left-linear connection
in the sense of our earlier considerations and additionally satisfies:
\begin{equation}
\nabla (m a ) = \nabla (m) a + \sigma ( m \tsa da ).
\end{equation}

Before we go on with calculating torsion and curvature in this
scheme we shall give our choice for $\sigma$. For the geometry of
triangulations we have a natural candidate
for $\sigma$. From the property \mbox{$\pi \circ (1 + \sigma) = 0$}
we immediately get the following:
\begin{equation}
\sigma \left( \link{p}{q} \tsa \link{q}{r} \right)
= \left\{
\begin{array}{ll}  -1 & \hbox{if\ } p,r,q  \hbox{\ form a triangle} \\
 1 & \hbox{otherwise} \end{array} \right\}.
\end{equation}
Of course, we could have chosen some other constant, not necessarily $1$,
in the second case, however, it is quite convenient to have normalization
$\sigma^2 \equiv \hbox{id}$. By no means is this choice of $\sigma$
unique, however the study of all possible choices would be an impossible
task, besides as we try to keep our considerations as close to the classical
case as possible, we use the argument that the above chosen $\sigma$ is
a generic for all triangulations and therefore we were justified in our
selection.

An arbitrary bimodule connection has the form:
\begin{equation}
\nabla ( \link{p}{q} ) =
\sum_s \link{s}{p} \tsa \link{p}{q}
- \sum_r \sigma \left( \link{p}{q} \tsa \link{q}{r} \right)
+ \sum_r \Gamma^{pq}_{rq} \link{p}{r} \tsa \link{r}{q}.
\end{equation}

The torsion constraint is just as it was in the case of
left-linear connections:
\begin{equation}
\Gamma^{pq}_{rq} = -1, \hbox{ for $p,q,r$ forming a triangle}
\label{Bi-1} \end{equation}

The metric compatibility condition also remains the same:
\begin{equation}
\Gamma^{pq}_{sq} g_{sq} = \overline{\Gamma}^{sq}_{pq} g_{pq},
\end{equation}
and by using it and (\ref{Bi-1}) we again reach the conclusion
that all $r_{pq}$ are equal to each other.

One of the nice properties of bimodule connections is that
they could be extended to the tensor product of forms, for
$\omega_1 \tsa \omega_2 \in \Om{1} \tsa \Om{1}$ we define
the following:
\begin{equation}
\nabla' ( \omega_1 \tsa \omega_2) = \nabla(\omega_1) \tsa \omega_2
+ (\sigma \tsa \hbox{id}) \left( \omega_1 \tsa ( \nabla \omega_2) \right).
\label{nabla2}
\end{equation}

In the discussion of bimodule connections for some other geometries
\cite{MAD},  it was proposed that the metric should satisfy an additional
symmetry relation of the form:
\begin{equation}
(g \tsa \hbox{id}) \circ \nabla' = d \circ g = (\hbox{id} \tsa g) \circ
\nabla', \label{susym}
\end{equation}
where $\nabla'$ is as in (\ref{nabla2}). Surprisingly, it
appears that neither equation could be satisfied in our triangulation
geometry and a simple counter-example to (\ref{susym}) can be constructed
using $p,q,s$ such that they form a triangle.

Finally we may proceed and calculate the curvature tensor, Ricci
tensor and the scalar of curvature. To make the formula simpler we
introduce the following notation, let $\delta_{pqr}$ be $1$ if
$p,r,q$ form a triangle and $0$ otherwise. Then $\eta_{prq}$ would
be $1 - 2 \delta_{pqr}$. Using this notation we rewrite
the curvature $R=\nabla^2$ as the following map:

\begin{eqnarray*}
R \left( \link{p}{q} \right) & = &
\sum_{r,s \not= q} \left(  \eta_{pqs} - \eta_{pqs} \delta_{qrs} \right)
 \dlink{p}{q}{r} \tsa \link{r}{s} \\
&& \sum_{r,s} \left( - \eta_{pqs} + \eta_{qrs} \right)
 \dlink{p}{r}{q} \tsa \link{q}{s} \\
&& \sum_{r,s} \left( \delta_{psq} - \delta_{pqr}\delta_{qrs}
+\delta_{pqs} \delta_{rq} \right)
 \dlink{p}{r}{s} \tsa \link{s}{q} \\
\end{eqnarray*}

Though $\nabla$ respected the bimodule structure, $R$ is not a bimodule
map, it is, however, completely determined by the geometric structure
of the simplicial manifold. Using the same procedure as in the case
of left-linear connections we get:

\begin{equation}
Ric = \sum_{p,r,s} \link{q}{r} \tsa \link{r}{s}
\left( \delta_{pqr} \eta_{pqs} - \delta_{pqr} \eta_{pqs} \delta_{qrs}
+ \delta_{pqr} \delta_{sq} \right)
\end{equation}

The scalar of curvature function becomes:

\begin{equation}
R_q = \sum_{p,r} \mu_r \delta_{pqr} .
\end{equation}

and the action would have the following form:

\begin{equation}
S =  \sum_{p,r,q} \mu_r \mu_q \delta_{prq}
\end{equation}

This could be easily rewritten in a more symmetric way:

\begin{equation}
S = \sum_{\triangle} \mu(\triangle)
\end{equation}
where
$$ \mu(\triangle) = \mu_p\mu_q\mu_r \left( \frac{1}{\mu_p} + \frac{1}{\mu_q}
+ \frac{1}{\mu_r} \right). $$

This action also seems to be an interesting candidate for further
investigations, with two possibilities opening. First, we may treat
$\mu$ as a dynamical field and investigate it on a fixed simplicial
manifold. The other choice is to fix the measure and change the
geometry, similarly as in the random triangulation simulations
of quantum gravity.

One good candidate for a measure is $\mu_p = \frac{n_p}{N(n+1)}$,
where $n_p$ denotes the number of $n$-dimensional simplices at
point $p$, $n$ is the dimension of the simplicial manifold. and $N$
is the total number of $n$-dimensional simplices. Clearly,
$$\int 1 = \sum_p \mu_p = \sum_p \frac{n_p}{N(n+1)} = 1,  $$
as the integration just counts (locally) $n$-dimensional simplices.

A further investigation of the proposed action and the properties
of the resulting model would certainly prove worthwhile. Especially
the most interesting would be the critical behavior of the determined
model and its relations with the models of discrete gravity constructed
and tested up to now.

\section{Conclusions}

The noncommutative differential geometry is a powerful tool, which has
we have used to study the simplicial manifold from the differential
geometry point of view.

Our main task was to investigate the linear connections for this models
in hope that we might obtain a satisfactory action candidate for simplicial
gravity. Comparing the three possible approaches we reach the conclusion
that neither of the proposed definitions for linear connections is flawless.
Only for strictly {\em bimodule} objects, i.e. bimodule connections we were
able
to get the action, which depends only on the {\em metric} in the general sense.
This actions are definitely worth further investigations, in particular it
would be
interesting to find what is their relation with the gravity actions obtained as
a
discretized approximation of the continuous Hilbert-Einstein action.

We have shown also that the metric, understood as a certain bimodule map
with some additional properties, agrees with picture of the metric arising
from the Dirac operator, indeed in both approaches we were able to sort
out of the metric its {\em distance} and its {\em measure} part. Note, that
the additional symmetry constraint on the metric, as proposed by some
authors cannot be introduce in case of discrete geometry.

Simplicial manifolds are not only a nice model of discretized geometry,
relevant for testing physical models but they might be a candidate for
the model of space-time structure at Plank scale lengths. Therefore the
investigation of gravity models and their geometry is important not only
for testing various definitions of linear connections but also for our
understanding of fundamental structure of space-time.

\ \\
{\bf Acknowledgments:} The author would like to thank
Mario Paschke for comments and remarks.
%%%%%%%%%%%%%%%%%%%%%%%%%%%%%%%%%
%%%% Appendix & Bibliography
%%%%%%%%%%%%%%%%%%%%%%%%%%%%%%%%%
\section{Appendix - Differential algebra on simplicial manifolds}

We present here the precise mathematical description of the differential
algebra on a simplicial manifold. As a first step we remind the
construction of the universal differential envelope of the algebra
$\CA$ of complex valued functions on the lattice made of vertices $M$.

Let $\Omega^n\CA$ be the vector space of maps from $M^{n+1} \to \C$,
i.e. functions of $n+1$ variables on $M$, which vanish on any subset
of $M^{n+1}$ containing a two-dimensional diagonal:
$$ \hbox{diag}(M \times M) = \{m,m\} \in M \times M, \;\;\; m \in M.$$
Of course, $\Omega^0\CA = \CA$. Then let $\Omega\CA$ be a direct sum of
all $\Omega^n\CA$ from $n=0$ to $n=\infty$ (an inductive limit).
One introduces the following multiplication law between elements of
$\Omega\CA$, for $f \in \Omega^p\CA$ and $g \in \Omega^w\CA$ we
define $fg \in \Omega^{p+w}\CA$ to be the following function:
%%%%%%
$$ (fg) (a_0, \ldots a_{p+w}) = f(a_0,\ldots,a_p) g(a_p, \ldots, a_{p+w}). $$
%%%%%%
It remains to verify that $fg$ satisfies the additional condition. Indeed,
if we assume that for some $0{\leq}i{\leq}p+w$ we have $a_i = a_{i+1}$
(which means that then the arguments belong to a subset of $M^{p+w+1}$
containing the two-dimensional diagonal), we would have either $f$ vanishing
(for $i{\leq}p-1$) or $g$ vanishing (for $i{\geq}p+1$).

Of course, the above product is associative and non-commutative (apart from
the case when $f$ and $g$ are both from $\CA$).
The differential structure on $\Omega\CA$ is given by the external
derivative operator $d$, for $f \in \Omega^p\CA$ we define $df \in
\Omega^{p+1}$:
\begin{equation}
df(a_0, \ldots, a_{p+1}) = \sum_{i=0}^{p+1} (-1)^i f (a_0, \ldots,
\hat{a}_i, \ldots, a_{p+1}),
\label{def-d}
\end{equation}
where $\hat{a_i}$ denotes that we throw away the $i$-th entry (since $f$
is a function of only $p+1$ arguments).
The operator $d$ satisfies the graded Leibniz rule, for $f \in \Omega^p\CA$
and $g \in \Omega^w\CA$ we have:
$$ d(fg) = (df)g + (-1)^p f (dg), $$
the proof of which we present below:
%%%%%%%%%%
\begin{eqnarray*}
& d( fg) (a_0, \ldots , a_{p+w+1}) = \sum_{i=0}^{p+w+1} (-1)^i (fg)
(a_0, \ldots , \hat{a}_{i}, \ldots, a_{p+w+1}) = & \\
& = \sum_{i=0}^{p}  (-1)^i  f (a_0, \ldots , \hat{a}_{i}, \ldots, a_{p+1})
g(a_{p+1}, \ldots, a_{p+w+1}) + & \\
& + \sum_{i=0}^{w+1} (-1)^{i+p+1} f(a_0,\ldots,a_p) g(a_p, \ldots,
\hat{a}_{i+p}, \ldots, a_{p+w+1}) = \ldots &
\end{eqnarray*}
%%%%%%%%%%
to recover the desired expression we must add to the last component
of the sum the following term:
$$(-1)^p f(a_0,\ldots,a_p)g(a_{p+1},\ldots, a_{p+w+1})$$
and its opposite,  $(-1)^{p+1} f(a_0,\ldots,a_p)g(a_{p+1},\ldots, a_{p+w+1})$
(so that they cancel together) to the first component. Then, grouping the
elements in the sums together we obtain:
$$ \ldots = (df) g + (-1)^p f (dg). $$
The usual complex conjugation on $\CA$ extends to the universal algebra
$\Omega\CA$ if we define, for $f \in \Omega^p\CA$:
$$ (f^\star)(a_0,a_1\ldots,a_{p-1}, a_p)  =  (-1)^p \bar{f}(a_p,a_{p-1},
\ldots, a_1, a_0), $$
so that $(df)^\star = (-1)^p d(f^\star)$ for a form $f$ of degree $p$.

The universal differential algebra is, however, too big to describe the
simplicial geometry. In fact, it makes use only of $0$-dimensional
structures, vertices, and does not rely on the higher-dimensional
objects. It is useful for the practical purposes coming from universality
property. In the case of the simplicial manifold it is  easier to
construct an universal object and then proceed with some quotient
construction. We shall apply this procedure to the simplicial geometry.

\subsection{Simplicial ideals.}

Let us consider the subset of $\Omega^p\CA$ consisting of all
functions, which vanish on all points $\{a_0,\ldots,a_p\} \in M^{p+1}$
such that all points are different from each other and belong to one
simplex of dimension $p$. Of course, such functions form a vector subspace of
$\Omega^p\CA$ and, moreover, this subspace remains invariant under
multiplications by elements of $\CA$ from both sides, hence it is
a subbimodule of $\Omega^p\CA$ over $\CA$, which we will denote
as $\CM^p$. Clearly, $\CM^0 = 0$, for $k$ greater then the dimension
of the manifold we  set $\CM^k = \Omega^k\CA$.
We shall now  demonstrate that $d\CM^p \subset \CM^{p+1}$.
Let us take $f \in \CM^p$ and evaluate $df$ on points $\{a_0,\ldots,a_{p+1}\}$.
Due to the definition (\ref{def-d}) it is sufficient to check that each
component of the sum vanishes, i.e.:
%%%%%%%%%%
$$  f(a_0, \ldots, a_{i-1},a_{i+1},\ldots, a_{p+1}) = 0. $$
%%%%%%%%%%
This is true, as all points $\{a_0,\ldots,a_{i-1},a_{i+1},\ldots,a_{p+1}\}$ are
different from each other by assumption. From the construction of simplices
we know that if $\{a_0,\ldots,a_{p}\}$ belongs to a $p+1$-dimensional
simplex then $\{a_0,\ldots, a_{i-1},a_{i+1},\ldots,a_{p}\}$ belong to one
of its wall's, so to a $p$ dimensional simplex. Since $f$ belongs to $\CM^p$
the expression above vanishes and indeed $df \in \CM^{p+1}$. It remains to
verify that the product of two forms $f$ and $g$ of degrees $p$ and $w$,
respectively, lies in $\CM^{p+w}$ whenever one of them is in the $\CM^i$.
Taking $p+w+1$ points, different from each other and belonging to a $p+w$
dimensional simplex, we know that the first $p+1$ of them (or the last $w+1$)
are again different from each other and belong to a $p$ dimensional (or
$w$-dimensional, respectively) subsimplex. Therefore the product of $f$ and
$g$ will vanish provided that either $f$ or $g$ are in $\CM^i$. Using
similar simple arguments we observe that $(\CM^i)^\star = \CM^i$.

The last observation finishes our construction as we have shown that the
direct sum:
%%%%%%%%%
$$ \CI = \oplus_{i=0}^\infty \CM^i, $$
%%%%%%%%%
is a differential (star) ideal of the algebra $\Omega\CA$, i.e., $d\CI
\subset \CI$. We can construct the quotient algebra:
$$ \Omega_s(\CA) = \Omega/\CI, $$
which will have a well-defined star and differential structure. The
projection $\pi : \Omega\CA \to \Omega_s(\CA)$ is a morphism of
differential algebras. Below we state some important properties of
$\Omega_s(\CA)$:
\begin{itemize}
\item $\Omega^0_s(\CA) = \CA$
\item $\Omega_s(\CA)$ is finite dimensional, $\Omega_s^k(\CA)=0$ for
      $k$ greater than the dimension of the simplicial manifold.
\item There exist a natural embedding: $i : \Omega_s(\CA) \hookrightarrow
      \Omega\CA$, constructed in the following way. If
      $\omega \in \Omega^p_S(\CA)$ we take $i(\omega)$ to be defined as
      the following function:
%%%%
$$ i(\omega)(a_0,\ldots,a_p) = \left\{
\begin{array}{ll}
F(a_0,\ldots,a_p) & \hbox{if\ } \{ a_0,\ldots,a_p \} \hbox{\ are vertices
of a $p$-dimensional} \\
&  \hbox{simplex, different from each other, and\ } F \in \pi^{-1}(\omega). \\
0 & \hbox{otherwise}
\end{array} \right.
$$
%%%%
\item There exists a graded trace, i.e., a map $\Omega^n_s(\CA) \to \C$,
      such that $\int df = 0$.
\end{itemize}

The first two observations follow directly from the construction, so we
concentrate on the remaining two. It is easy to see that the definition
of $i(\omega)$ does not depend on the choice of $F \in \pi^{-1}(\omega)$,
so that $i$ is a well-defined map. Of course, $\pi \circ i = \id$, however,
clearly it is neither an algebra nor a differential map ($d \circ i \not=
i \circ d$). However, if we introduce a multiplication rule on the
image of $i$ just as the image of the multiplication in the quotient:
$$ i(\omega) i(\rho) := i(\omega\rho), $$
we easily obtain an algebra structure on $i(\Omega_S(\CA))$. This product
is quite simple and we shall discuss it in a more detailed way for the
forms which generate $\Omega_s(\CA)$. In the same
way we could transport the external derivative $d$, defining $d(i(f)) = i(df)$.

{}From now on, we shall very often work with the image of a differential form
under the map $i$ rather than with the element of the quotient, using the
image of multiplication and $d$.

The construction of the graded trace is as follows. First, define the map
$\int: \Omega^n\CA \to \C$:

$$ \int f = \sum_{ \hbox{\tiny A}} \sum_{\sigma} (-1)^\sigma
            i(f)\left( a_{\sigma(0)},\ldots,a_{\sigma(n)} \right), $$
where $\sigma$ denotes the permutation of $n+1$ elements and the first sum
is over all $n$-dimensional simplices.

We shall prove now that the above defined operation has all properties of
a graded trace. Clearly, it is linear and $\int f^* = \overline{\int f}$.
Calculate $\int df$:

\begin{eqnarray*}
&& \int df = \sum_{ \hbox{\tiny A}} \sum_{\sigma} (-1)^\sigma
  i(df) \left( a_{\sigma(0)},\ldots,a_{\sigma(n)} \right) =  \\
&& = \sum_{ \hbox{\tiny A}} \sum_{\sigma} (-1)^\sigma
\sum_{i=0}^n (-1)^i f \left( a_{\sigma(0)},\ldots, a_{\sigma(i-1)},
a_{\sigma(i+1)}, \ldots, a_{\sigma(n)} \right) = \ldots
\end{eqnarray*}
Now, let us choose among all simplices $A$ and $B$, such that
they are glued along an $n-1$-dimension simplex
$W =\{w_1,\ldots,w_{n-1}\}$. We shall now investigate all elements of
the sum, which have the form  $f(w_\tau(0),\ldots,w_{\tau(n-1)})$ for a
certain permutation $\tau$ of $n-1$ elements. Obviously, one contribution
comes from simplex $A$ and another one from $B$:
%%%%%%%%%%%%%%%%
$$ \sum_i \sum_{\sigma_A}' (-1)^{\sigma_A}
(-1)^i f\left( w_{\tau(0)},\ldots,w_{\tau(n-1)} \right)
+ \sum_i \sum_{\sigma_B}'  (-1)^{\sigma_A}
(-1)^i f\left( w_{\tau(0)},\ldots,w_{\tau(n-1)} \right),$$
%%%%%%%%%%%%%%%
where the prime over sum indicate that we are summing only over such
$\sigma_A$ that satisfy: (for a given $i$):
$$ \{  a_{\sigma_A(0)},\ldots, a_{\sigma_A(i-1)},
a_{\sigma_A(i+1)}, \ldots, a_{\sigma_A(n)} \}
\equiv \{ w_\tau(0),\ldots,w_{\tau(n-1)} \} $$
with the same property, respectively for the sum over $\sigma_B$.

Now we can calculate the sums. For the simplex $A$ the signs
$(-1)^{\sigma_A}$ and $(-1)^i$ give together a sign, which
corresponds to the orientation of $\tau(W)$ induced by $A$, so after
summing all elements we get $n$ multiplied by this sign. Similarly,
for the second sum we get the same but with the sign of orientation
of $W$ induced by $B$. However, since the manifold is orientable,
the orientation induced on $\tau(W)$ by $A$ is opposite to the one
induced by the orientation of $B$, hence this two sums add up to zero.
Because we can use this argument for all subsimplices $W$ of all
pairs $A$, $B$ this proves the desired property of the graded
trace.

\subsection{Differential algebra and links}

Before we complete our construction with some further quotient
construction (to eliminate overcounting of simplices) we shall
give here a relation between the above formal derivation of the
differential algebra and the picture of links.

We already know that the generators of the algebra $\CA$ could be
identified with points with the help of Kronecker delta functions:
$p(s) = \delta_{ps}$. We prove that this extends to higher order
forms.

Let $f$ be an element of $\Omega^p_S(\CA)$ and $Q$ be a $p$-dimensional
simplex $\{q_0,\ldots,q_p\}$. Then $i(f)$ could be written
in a unique way as:
\begin{equation}
i(f)  = \sum_Q \sum_\sigma C_\sigma^Q \;\; \delta_{\sigma(Q)},
\label{li-1}
\end{equation}
where $\sigma$ is a permutation of $p+1$ elements, $\C_\sigma^Q$
are complex numbers and $\delta_{\sigma(Q)}$ is the multiple
Kronecker delta function:
%%%%%%%%%%
$$ \delta_{\sigma(Q)} (b_0,\ldots,b_p) = \delta_{q_{\sigma(0)} b_0}
\delta_{a_{\sigma(1)}b_1}
\cdots \delta_{a_{\sigma(p-1)}b_{p-1}} \delta_{a_{\sigma(p)} b_p}. $$
%%%%%%%%%%

The proof is straightforward. First, by construction
we know that $i(f)$ vanishes on the set of arguments that are not
different vertices of a $p$-dimensional simplex. Let us fix a simplex $Q$,
and look at $i(f)$ on the subset $Q^{p+1}$. Then, the only non-vanishing
values of $i(f)$ are for permutations of different vertices. After summing
over all $p$-dimensional simplices $Q$ we obtain the formula (\ref{li-1}).

We see now that the functions $\delta_{\sigma(Q)}$ generate the image
of $\Omega_S(\CA)$. The multiplication between them, transported by $i$
from  $\Omega_S(\CA)$ is quite simple:
%%%%%%%%%
\begin{equation}
\delta_{\sigma(Q)} \delta_{\tau(P)} = \left\{
\begin{array}{ll}
\delta_{\sigma(Q)+\tau(P)} & \hbox{if $P \cup Q$ is a $p+q$-simplex} \\
0 & \hbox{otherwise}
\end{array} \right. \label{mu-1}
\end{equation}
%%%%%%%%%

We introduce now in a formal way a link. Let $l$ be the following
one-form, for any points $p$ and $q$ in a $1$-dimensional simplex,
we define the link one-form:
$$ l_p^q (b_0, b_1) = \delta_{p b_0} \delta_{q b_1}. $$
%%%%%%%%
We have used here the identification of $\Omega_S(\CA)$ with its image
under $i$ and, in fact, we have defined only $i(l^p_q)$, however, this
is sufficient as $i$ is injective. In our "pictorial"  notation:
%%%%%%%%
$$ l_p^k \; = \; \link{p}{q}. $$
%%%%%%%%

Next we shall use the result (\ref{li-1}) and the presentation of
one-forms by links and the rules for multiplication (\ref{mu-1}).
Since any multiple Kronecker delta function could be written as
products of functions for two entries (which are used to define
the link) we could rewrite (\ref{li-1}) using the links.

Then every $p$-form $f$ could be written as:
%%%%%%%%%%%%%%%%%%%%%%%%%
\begin{equation}
f = \sum_Q \sum_\sigma C_\sigma^Q \link{q_{\sigma(0)}}{q_{\sigma(1)}} \tsa
\cdots \tsa \link{a_{\sigma(p-1)}}{a_{\sigma(p)}}. \label{li-3}
\end{equation}
%%%%%%%%%%%%%%%%%%%%%%%%%
This follows directly from (\ref{li-1}) and the rules for product of
forms (\ref{mu-1}). Now, it is easy to verify the rules for the differentiation
of zero and one-forms, which we have presented earlier in the language of
links.
As an example we shall derive here formally the action of $d$ on a single
link.

Let $l_p^q$ be this link. First, let us take $i(l_p^q)$ being the Kronecker
delta function, then calculate $d$ of it, in the universal algebra
$\Omega\CA$:
%%%%%%%%%%%%%%%%%%%%%%%%
$$
\left( d \delta_{\{p,q\}} \right) ( b_0, b_1, b_2) =
\delta_p^{b_1} \delta_q^{b_2} - \delta_p^{b_0} \delta_q^{b_2}
+\delta_p^{b_0} \delta_q^{b_1}. $$
%%%%%%%%%%%%%%%%%%%%%%%%
Let us proceed the quotient operation. The nontrivial contributions
to $\pi(d \delta_{\{p,q\}})$ come only from such functions, which do
not vanish on $\{b_0,b_1,b_2\}$ being vertices of a triangle $T$.
Hence, we can write:
%%%%%%%%%%%%%%%%%%%%%%
$$
\pi\left( d \delta_{\{p,q\}} \right) =
\pi \left(
\sum_{r: \{r,p,q\} \in T}
\left(\delta_r^{b_0} \delta_p^{b_1} \delta_q^{b_2} -
 \delta_p^{b_0} \delta_r^{b_1} \delta_q^{b_2}
+\delta_p^{b_0} \delta_q^{b_1} \delta_r^{b_2}\right) \right) . $$
%%%%%%%%%%%%%%%%%%%%%%
Now, we observe that the components of the sum are actually in
the image of $i$ and each $\delta$ function corresponds to a
link. Using the previously introduced notation for the products
of two links we obtain:
%%%%%%%%%%%%%%%%%%%%%%
\begin{equation}
 d \link{p}{q}  \; = \; \sum_{r: \{r,p,q\} \in T}
\left( \dlink{r}{p}{q} + \dlink{p}{q}{r} - \dlink{p}{r}{q}
\right). \label{d-sin}
\end{equation}
%%%%%%%%%%%%%%%%%%%%%%

To finish our construction we have to add one more relation. Observe that
the vector space of $p$-forms has a dimension equal to $N_p
(p+1)!$, where $N_p$ is the number of $p$-simplices on our manifold. As
a bimodule is generated by $N_p (p-1)!$ elements. Therefore, only
for $p=1$ and $p=2$ the number of simplices corresponds to the number
of forms of the same dimension. For $p>2$ we could have different
$p$-forms with the same ''end-points''.

Our task would be to reduce the differential algebra in such a way that the
bimodule of $p$-forms is generated by $N_p$ elements. Hence, we would have
a one-to-one correspondence between $p$-simplices and $p$-forms. Since for
$p=1,2$ this is already the case, the relations would start on the level
of three-forms.

For every $p$-simplex $A = \{a_0,\ldots,a_p\}$ we introduce the set of
$p$-forms:
\begin{equation}
\CJ_A^p =  \link{a_0}{a_1} \tsa \cdots \link{a_{p-1}}{a_p} -
(-1)^\sigma  \link{a_0}{a_{\sigma(1)}} \tsa
\cdots \link{a_{\sigma(p-1)}}{a_p},
\label{ide-1}
\end{equation}
where $\sigma$ is a permutation of $p-1$ elements. Of
course, both components of the sum are different
(for $\sigma \not= \id$), so $\CJ_S$ is nonempty. Let $\CJ^p$
be the subbimodule of $\Omega^p_S(\CA)$ generated by all $\CJ_S^p$.
We shall prove that the direct sum of all $\CJ^p$ is a differential
ideal of $\Omega_S(\CA)$.

First, let us take $f \in \Omega^w_S(\CA)$ and $g \in \CJ^p$. Because
of linearity we might restrict ourselves to the case when $g \in\CJ_A^p$
and $f$ is just one of the generators (so that $i(f) = \delta_{\sigma{Q}}$
for some simplex $Q$ and permutation $\sigma$. The product, $fg$ will,
of course, vanish unless $\sigma(q_w) = a_0$ and $A \cup Q$ are a
$w+p$-dimensional simplex. In the latter case it is quite obvious that
the product would be of the form (\ref{ide-1}) and therefore in
$\CJ_{A \cup Q}^{w+p}$. So the direct sum of all $\CJ^p_S$ (summed both
for all $p=1,\ldots,n$ and all $p$-dimensional simplices $S$) would
be an ideal of $\Omega_S(\CA)$.

It remains to verify that $\CJ$ is a differential ideal, i.e. $d\CJ^p
\subset \CJ^{p+1}$. Again, by linearity, we need to prove it for $\CJ^p_A$,
for a simplex $A$. Let us calculate:
%%%%%%%%%%%%%%%%%%%%%%%%%
\begin{equation*}
d \left(  \link{a_0}{a_1} \tsa \cdots \link{a_{p-1}}{a_p} -
(-1)^\sigma  \link{a_0}{a_{\sigma(1)}} \tsa
\cdots \link{a_{\sigma(p-1)}}{a_p} \right) = \ldots
\end{equation*}
of course, to get the result we must use both the graded Leibniz rule
as well as the rule for the action of $d$ on a single link (\ref{d-sin}).
It is a simple exercise to see that the action of $d$ would give us a sum,
consisting of the above expression multiplied by a link at front of it,
at the end, or with one link in the middle split into the product of two
links (start and end-point remaining the same). Since the same applies
to each of the two components of the above sum, the general form of the
result is also the same, and therefore the result would lie in the sum
of $\sum_Q \CJ^{p+1}_Q$ for such $p+1$-simplices $Q$ that contain $A$.
Knowing that $\CJ$ is a differential ideal we may construct the quotient
$\Omega^*(\CA) = \Omega_S(\CA)/\CJ$, which is the differential algebra we
define  as the one corresponding to the geometry of the simplicial
manifold. Of course, the difference between $\Omega_S(\CA)$ and
$\Omega^*(\CA)$ starts only at the level of three forms, so in most physical
applications (which end at the level of two-forms) there is no necessity
to distinguish between them.
%%%%%%%%%%%%%%%%%%%%%%%%%%%%

%%%%%%%%%%%%%%%%%%%%%%%%%%%%%%%%%%%%%%%%%%%%%%%%
%%%%%%%%%%%%%%%%%%%%%%%%%%%%%%%%%%%%%%%%%%%%%%%%

\subsection{Linear representation and Dirac operator approach}

Quite often in noncommutative geometry the starting point of
the approach is not the differential structure, on which one build
metric and other structures but rather a representation of the
algebra on some Hilbert space and a Dirac operator, which gives
both the metric and differential structure. Here we should
briefly present the relevant construction for the simplicial
manifold.

Let ${\cal H}$ be the Hilbert space of dimension $\#{\cal A}$ (we
take the number of simplices to be finite), and let $\Psi_p$ denote
a vector in ${\cal H}$ associated with point $p$. Then there exist
a natural representation of the algebra ${\cal A}$ on the Hilbert space
given by:
$$ \pi(f)\Psi_p = f(p)\Psi_p. $$

Let $D$ denote an operator on ${\cal H}$, which does not commute with
the algebra and is self-adjoint. Then, we may define the differential
of the function as a commutator $i [D,f]$. Let's calculate it for
the generators of the algebra:
\begin{eqnarray}
[ D,p ] \Psi_q = D p \Psi_q - p D \Psi_q = \nonumber  \\
= \delta_{pq} D\Psi_q  - p D^w_q \Psi_w = \nonumber \\
= ( \delta_{pq} D^w_q - \delta_{pw} D^w_q ) \Psi_w \label{Di-1}
\end{eqnarray}

To verify that it corresponds to our previous formulae, let us
define a link operator $L_{pq}$ as the following isometry:
$$ L_{pq}\Psi_w = \delta_{wq} \Psi_p. $$
Then, we can rewrite (\ref{Di-1}) as:
\begin{equation}
\left( \sum_s D^s_p L_{sp} - \sum_s D^p_s L_{ps} \right) \Psi_q,
\end{equation}
which agrees with the abstract construction (the only difference is
that the links are rescaled). Of course, {\em a priori} the
operators $L_{pq}$ might be constructed for any pair of points $p,q$
and we must add the some information about the actual geometry of the
simplicial manifold - either by using only some of them (for $p,q$
connected by a link) or, preferably, assume that the Dirac operator $D$
has non-zero entries $D^p_q$ only and only if $p$ is connected with $q$.
The link operators $L_{pq}$ form are an extension of ${\cal A}$:
$$ L_{pq} L_{rs} = L_{ps} \delta_{qr} + s \delta_{qr} \delta_{ps}.$$
Equivalently, we may use $L_{pp} \equiv p$, then the above identity
becomes simpler $L_{pq} L_{rs} = L_{ps} \delta_{qr}$.

Since $D$ is self-adjoint and $L_{pq}^\dagger = L_{qp}$ we see that
$dp$ is, as an operator, self-adjoint.

Can we now recover the metric from the Dirac operator? Let us take two
points  $p$ and $q$, $p\not=q$, which simultaneously we may interpret
as operators on ${\cal H}$. Then, following the classical correspondence
we might look for the following identity:

\begin{equation}
\int g(dp,dq) = \hbox{Tr} \left( [D,p][D,q]\right).
\end{equation}

Of course, since our Hilbert space is finite dimensional we are using
the standard trace. Now, the left-hand side is as follows:

$$ \int g(dp,dq) = \mu_p g_{pq}  $$
whereas the right-hand side is equal:
$$ \hbox{Tr}  \sum_s \left( D^s_p D^q_r L_{sq} - D^p_s D^q_s L_{pq}
 + D^p_q D^q_r L_{pr} \right) = \ldots $$
Since $\hbox{Tr\ } L_{wz}$ is zero unless $w=z$ when it is $1$, we
obtain:
$$  \ldots  = 2 D^q_p D^p_q $$

Now we can see that having the Dirac operator $D$ we recover the metric
as it was defined earlier. Moreover, as the above expression is
symmetric in $p$ and $q$ (note that since $D$ is self-adjoint it must also
be real) we obtain the relation $\mu_p g_{pq} = \mu_q g_{qp}$, which we
have postulated.
%%%%%%%%%%%%%%%%%%%%%%%%%%%%%%%%%%%%%%%%%%%%%%%%%%%%%%%%%%%%%%%%

\end{document}